\documentclass[9pt,conference]{IEEEtran}
\usepackage[top=2.0cm, bottom=2.4cm, left=1.65cm, right=1.65cm]{geometry}
\usepackage{cite}
\usepackage{hyperref}
\usepackage{amsmath,amssymb,amsfonts,nccmath}
\usepackage{algpseudocode}
\usepackage{adjustbox}
\usepackage[linesnumbered,ruled,vlined]{algorithm2e}
\usepackage{multirow}
\usepackage[table,xcdraw]{xcolor}
\usepackage{textcomp}
\usepackage{xcolor}
\usepackage{graphicx}
\usepackage{subfigure}
\usepackage[bb=px]{mathalfa} 
\usepackage{tikz}
\usetikzlibrary{positioning, fit, arrows.meta, shapes}
\tikzstyle{line}=[draw]
\usepackage{enumerate}
\usepackage{longtable}
\usepackage{latexsym}
\usepackage{url}
\usepackage[switch]{lineno}
\usepackage[inline]{enumitem}
\usepackage{lipsum}
\usepackage{footnote}
\usepackage{caption}
\usepackage{subcaption}
\usepackage{rotating}
\usepackage{tabto}
\usepackage{comment}

\usepackage{fancyhdr}

\DeclareRobustCommand*{\IEEEauthorrefmark}[1]{%
  \raisebox{0pt}[0pt][0pt]{\textsuperscript{\footnotesize #1}}%
}

\def\BibTeX{{\rm B\kern-.05em{\sc i\kern-.025em b}\kern-.08em
    T\kern-.1667em\lower.7ex\hbox{E}\kern-.125emX}}
\begin{document}

\title{X-CBA: Explainability Aided \textbf{C}at\textbf{B}oosted \textbf{A}nomal-E for Intrusion Detection System}

\author{%
  \IEEEauthorblockN{%
    Kiymet Kaya \IEEEauthorrefmark{1},
    Elif Ak \IEEEauthorrefmark{1,3},
    Sumeyye Bas \IEEEauthorrefmark{2,3},
    Berk Canberk \IEEEauthorrefmark{2,4},
    Sule Gunduz Oguducu\IEEEauthorrefmark{2}
  }%
  \IEEEauthorblockA{\IEEEauthorrefmark{1} Istanbul Technical University, Department of Computer Engineering, Istanbul, Turkey }%
  \IEEEauthorblockA{\IEEEauthorrefmark{2} Istanbul Technical University, Department of Artificial Intelligence and Data Engineering, Istanbul, Turkey }%
  \IEEEauthorblockA{\IEEEauthorrefmark{3} BTS Group, Istanbul, Turkey}
\IEEEauthorblockA{\IEEEauthorrefmark{4} School of Computing, Engineering and Built Environment, Edinburgh Napier University, Edinburgh}
Email: \{kayak16, akeli, bass20\}@itu.edu.tr, B.Canberk@napier.ac.uk, sgunduz@itu.edu.tr}

\maketitle

\thispagestyle{fancy}   
\fancyhead{}                
\lhead{Accepted by 2024 IEEE International Conference on Communications (ICC), \copyright2023 IEEE}
\cfoot{}
\renewcommand{\headrulewidth}{0pt}

\begingroup\renewcommand\thefootnote{\textsection}
\endgroup
\begin{abstract}
The effectiveness of Intrusion Detection Systems (IDS) is critical in an era where cyber threats are becoming increasingly complex. Machine learning (ML) and deep learning (DL) models provide an efficient and accurate solution for identifying attacks and anomalies in computer networks. However, using ML and DL models in IDS has led to a trust deficit due to their non-transparent decision-making. This transparency gap in IDS research is significant, affecting confidence and accountability. To address, this paper introduces a novel Explainable IDS approach, called X-CBA, that leverages the structural advantages of Graph Neural Networks (GNNs) to effectively process network traffic data, while also adapting a new Explainable AI (XAI) methodology. Unlike most GNN-based IDS that depend on labeled network traffic and node features, thereby overlooking critical packet-level information, our approach leverages a broader range of traffic data through \textit{network flows}, including edge attributes, to improve detection capabilities and adapt to novel threats. Through empirical testing, we establish that our approach not only achieves high accuracy with 99.47\% in threat detection but also advances the field by providing clear, actionable explanations of its analytical outcomes. This research also aims to bridge the current gap and facilitate the broader integration of ML/DL technologies in cybersecurity defenses by offering a local and global explainability solution that is both precise and interpretable.

\end{abstract}


\section{Introduction} \label{sec:1}
In the contemporary digital environment, the continued increase in sophisticated cyber threats still leaves security mechanisms inadequate. Gartner forecasts that by 2024, a minimum of 50\% of organizations will adopt Machine Learning (ML) or Deep Learning (DL) aided Security Operations Centers (SoCs) for faster cyberattack detection, a shift that is already underway with substantial investments from leading firms in AI for enhanced security. However, it is also well-known that there is a notable hesitance among enterprises to adopt ML/DL-augmented network Intrusion Detection Systems (IDS) due to the inconceivable \textit{black-box} decision-making processes, which are perceived as complicated, unpredictable, and unreliable \cite{9927396}. Addressing these concerns, the current IDS literature mainly prioritizes (i) developing advanced ML/DL models for sophisticated attack detection \cite{10005100, 9322153}, and (ii) employing explainable AI to demystify ML/DL decision-making \cite{9927396}. For the first aspect, recent studies indicate that Graph Neural Networks (GNNs), a subset of DL models, are particularly promising for IDS \cite{10123384}. Since the natural structure of network IDS is graph-based, where nodes are the network devices (e.g. routers, hosts, etc. ) and edges are connections, packet transfers, or network flows between network devices. In this way, GNNs reveal the impact of malicious activities on the network's topology and leverage the neighboring information among network entities for improved detection. Moreover, the current studies show that using \textit{network flows} rather than individually \textit{packet-based} monitoring is more suitable in IDS studies \cite{10005100}, since aggregating network features helps to reveal diverse and heterogeneous characteristics of cyberattacks.  Expanding on the second aspect, there is a concerted academic effort to incorporate Explainable AI (XAI) models into various ML/DL frameworks to provide local and global explanations of their operations. This endeavor aims to exploit the rationale behind model predictions, whether by clarifying the significance of specific data points (local explainability) or by shedding light on the model's overall behavior (global explainability) \cite{9927396}. Considering the research efforts on IDS and the security needs of organizations, it is evident that an advanced GNN-based methodology, combined with an appropriate XAI framework, would bridge the existing divide. Nevertheless, surprisingly, there is a notable scarcity in the literature, with few studies, such as the one by Baahmed et al. \cite{baahmedusing}, implementing an advanced GNN model alongside an XAI tool. To fill this gap, the proposed X-CBA enhances the attack detection results of state-of-the-art IDS, and it demonstrates that the PGExplainer \cite{luo2020parameterized} provides superior performance in explaining the operational dynamics of sophisticated GNNs for \textit{network flows} using local and global explainability. The main contributions of our study can be summarized as follows:

\begin{itemize}
    \item \textbf{Flow-based Network Data and Graph Edge Embeddings}: The proposed intrusion detection system (IDS) approach uses \textit{network flows} with many critical network metrics, representing them as graph \textit{edge embeddings}. Using flow-based network data with well-modeled graph embedding is particularly effective in detecting a range of cyber threats, including BruteForce, DDoS, DoS, and sophisticated attacks like Bot, by uncovering distinctive threat patterns.
    
    \item \textbf{Enhanced Detection Performance with GNN and CatBoost Integration:} Our novel method integrates a GNN-based detection pipeline with the CatBoost classifier, achieving higher accuracy, F1 score, and detection rates compared to existing state-of-the-art solutions in intrusion detection.
    
    \item \textbf{Advanced Explainability with PGExplainer Implementation:} The system implements PGExplainer, offering both local and global insights into the decision-making processes of the GNN-based IDS. This approach outperforms baseline explainability models, particularly due to its ability to operate in inductive settings, offer a non-black-box evaluation approach, and explain multiple instances collectively, making it highly suitable for flow-based IDS network data.
\end{itemize}

The rest of the paper is organized as follows. Section \ref{sec:lit} presents related works.Section \ref{sec:met} gives details of our proposed model and background methodology. Section \ref{sec:eval} presents baseline models and experimental results. Lastly, Section \ref{sec:conc} concludes the paper.

\begin{figure*}[t]
\centering
\includegraphics[width=.9\linewidth]{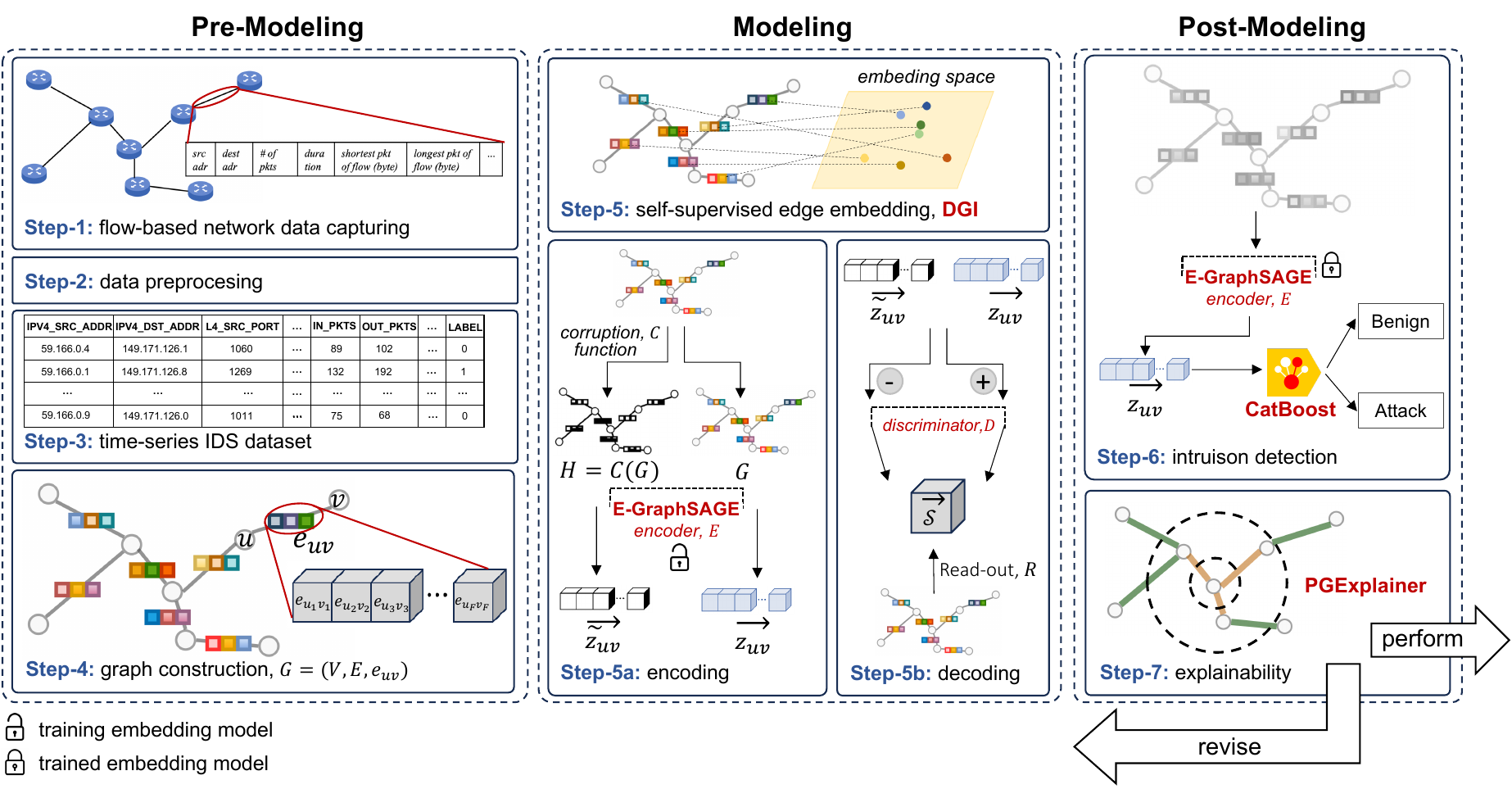}
\caption{X-CBA: Explainability Aided CatBoosted Anomal-E Intrusion Detection System based on the DARPA \cite{gunning2019darpa} Recommendation}
\label{fig_proposed}
\end{figure*}

\section{Literature Review} \label{sec:lit}

The benefit of GNNs in intrusion detection is that they take into account the properties of the computer network topology such as relationships between nodes. However, this also increases the complexity of the prediction model and can be computationally costly in large graphs and complex network topologies \cite{du2019graph, 8292772}. Computational speed is important, especially in large networks and real-time applications like intrusion detection. For this reason, many of the current works \cite{caville2022anomal} utilize a less complex GNN model for representation learning and predicting network anomalies mainly with tree-based ensemble models. Moreover, there are still some works \cite{fraihat2023intrusion, 9525087} that only use tree-based eXtreme Gradient Boosting (XGBoost) and CatBoost methods for forecasting, since it is hard to learn a powerful GNN when computing resources are limited \cite{du2019graph}. 

Among the studies on intrusion detection on NSL-KDD data, in \cite{mane2021explaining} the prediction model 3-layers neural network (NN) was explained with SHAP and LIME, in \cite{wang2020explainable} anomalies were predicted with RF and the explanations were conducted with SHAP. Patil et al. performed intrusion detection with Random Forest models and used LIME for model explanation \cite{patil2022explainable}. When reviewing the literature on explainability in intrusion detection models, most studies favor non-GNN-based explanation methods like SHAP and LIME, despite employing GNN-based prediction models. Only a few studies \cite{baahmedusing, lo2023xg} include GNN-based models and explanations. In both of these studies, the authors only observed the performance of GNNExplainer to find the important components of the network whereas comparative results in terms of XAI models were not observed. To this end, this study focuses on improving the predictive accuracy of intrusion detection models and conducting a comparative analysis of XAI methods on computer network data, that provide additional insights to understand and interpret conclusions of IDS.

\section{Methodology} \label{sec:met}
The overview of the Methodology is organized in two steps. In the first step, \textit{Background Methods} E-GraphSAGE, DGI, CatBoost, and PGExplainer shown in red-text in Fig. \ref{fig_proposed} are presented to make the proposed model easy to understand. Secondly, the flow of the X-CBA which is presented in Fig. \ref{fig_proposed}.

\subsection{Background Methods}

\subsubsection{\textbf{Edge Embedding: E-GraphSAGE}}

E-GraphSAGE \cite{lo2022graphsage} algorithm requires a graph input $G(\mathbb{V}, \mathbb{E})$ consisting of nodes ($\mathbb{V}$) and edges ($\mathbb{E}$). As an extended version of the GraphSAGE algorithm, it allows the use of edge features (\{$\textbf{e}_{uv}$ $\forall uv \in \mathbb{E}$\}) between each node $u$ and node $v$ in addition to node features ($~\overrightarrow{x_v}$) for message propagation. By utilising the edge attributes and graph topology, E-GraphSage generates output containing a new vectorial representation a.k.a. embeddings for each node ($\textbf{z}_{v}^K$) and each edge ($\textbf{z}_{uv}^K$). The flow-based NIDS datasets only consist of flow (edge) features rather than node features. Therefore, the feature vectors of nodes are set as $~\overrightarrow{x_v}$ = \{1, . . . , 1\} and the dimension of all one constant vector is the same as the number of edge features. E-GraphSAGE samples a fixed number of neighbors $k$ for each node in the graph data and aggregates information from the sampled neighbors to create an embedding for the destination node and edge embeddings. The initial value for $k$=0 ($\textbf{h}_v^{0}$) is the feature vector ($~\overrightarrow{x_v}$) of that node for all nodes.

\begin{equation}
\begin{aligned}
\label{esage_2}
    \textbf{h}_{N(v)}^k \leftarrow AGG_k\left(\{\textbf{h}_u^{k-1} \| e_{uv}^{k-1}, \forall u \in N(v), uv \in \mathbb{E}\}\right); \\
     \textbf{h}_v^k \leftarrow \sigma\left(W^k \cdot \text{CONCAT}(\textbf{h}_v^{k-1}, \textbf{h}_{N(v)}^k)\right)
 \end{aligned}
\end{equation}

In each iteration from $k$=1 to $K$, for all nodes in the node-set $\mathbb{V}$, the node $v$'s neighborhood is initially sampled and the information from the sampled nodes is collected into a single vector. Next, as in \ref{esage_2}, for all $k$ and $v$ values, the aggregated information $\textbf{h}_{N(v)}^k$ at the $k$-th layer and at node $v$, based of the sampled neighborhood $N(v)$ is calculated with the help of neighborhood aggregator function $AGG_k$. Here, $e_{uv}^{k-1}$ are the features of edge $uv$ from $N(v)$, the sampled neighborhood of node $v$, at layer $k$-1.

\begin{equation}
\label{esage_3}
    \textbf{z}_{uv}^K \leftarrow \text{CONCAT}(\textbf{z}_u^K, \textbf{z}_v^K)
\end{equation}

The aggregated embeddings of the sampled neighborhood $\textbf{h}_{N(v)}^k$ are then concatenated with the node’s embedding from the previous layer $\textbf{h}_v^{k-1}$. Here, the critical difference from the GraphSAGE is having the edge features. The final node embeddings at depth $K$ are assigned and the edge embeddings $\textbf{z}_{uv}^K$ for each edge $uv$ are calculated as the concatenation of the node embeddings of nodes $u$ and $v$ as in \ref{esage_3}.

\subsubsection{\textbf{Self-supervised Learning: Deep Graph Infomax}}

Self-supervised learning aims to learn the underlying features of the data by creating its own pseudo-labels from unlabeled data. The pseudo-labels are the labels automatically obtained by the self-supervised model, not the ground truths of the data. Teaching these pseudo-labels contributes to the creation of good representations for the data and can improve the prediction performance of the supervised learning model to be used later. DGI \cite{velivckovic2018deep} provides self-supervised learning by maximizing mutual local-global information and the trained Encoder of DGI can be reused to generate edge/node embeddings for subsequent tasks such as edge/node classification. The details of the DGI model presented in Step 5 of Fig. \ref{fig_proposed} are as follows:

\begin{itemize}

    \item A corruption function $C$ (i.e., a random permutation of the input graph node features which add or remove nodes from the adjacency matrix $A$) is used to generate a negative (corrupted) graph representation $H=C(G)$ from the input graph ($G$).
    
    \item An encoder $E$, which can be any existing GNN such as E-GraphSAGE, generates edge embeddings both for the input graph ($G$) and corrupted graph ($H$).

    \item The readout function $R$, which is at the core of DGI, aggregates edge embeddings by taking average of them and then processes them through a sigmoid function to calculate a global graph summary $~\overrightarrow{s}$ (a single embed vector of the entire graph). 

    \item The discriminator $D$ then assesses these edge embeddings (a real edge embedding \textbf{$~\overrightarrow{\textbf{z}_{uv(i)}}$} and a corrupted edge embedding \textbf{$~\overrightarrow{\widetilde{\textbf{z}}_{uv(j)}}$} ) using the global summary $~\overrightarrow{s}$ as a guide. 

    \begin{equation}
    \begin{aligned}
    \label{eq_dgi}
    L = \frac{1}{P+S} (\sum_{i=1}^{P}E_{(X,A))}[log D (~\overrightarrow{\textbf{z}_{uv(i)}}, ~\overrightarrow{s})] \\
    + \sum_{j=1}^{S}E_{(\overline{X},\overline{A}))}[(1- log D (~\overrightarrow{\widetilde{\textbf{z}}_{uv(j)}}, ~\overrightarrow{s}))])
    \end{aligned}
    \end{equation}
    
    \item Comparisons by the discriminator $D$ provides a score between 0 and 1, with the help of binary cross-entropy loss objective function in \ref{eq_dgi} to discriminate the embedding of the real edge and the corrupted edge to train the encoder $E$.
    
\end{itemize}

\subsubsection{\textbf{Intrusion Detection: Catboost Classifier}} CatBoostClassifier is a gradient-boosting ML algorithm known for its high predictive accuracy and speed. It employs techniques such as ordered boosting and oblivious trees to handle various data types effectively and mitigate overfitting. 

\subsubsection{\textbf{Explainability: PGExplainer}}
PGExplainer \cite{luo2020parameterized} offers explanations on a global level across numerous instances by developing a shared explanation network from nodes and graph representations of the GNN model. It seeks to locate a crucial subgraph that includes the most important nodes for the predictions of the given trained GNN model makes predictions by removing nodes and attributes, and analyzes their effects on the output of the GNN model. Removing nodes also means removing the edges that are the endpoints of that node from the graph, which leads to the identification of crucial pathways. PGExplainer divides the input graph $G$ into two subgraphs as in \ref{pge_1}, $G_S$ represents the crucial subgraph and $\Delta G$ is comprised of unnecessary edges.

\begin{equation}
\resizebox{0.25\hsize}{!}{
$G = G_S  + \Delta G$
}
\label{pge_1}
\end{equation}

\begin{equation}
\resizebox{0.7\hsize}{!}{
    $\max_{G_S}MI(Y, G_S) = H(Y) - H(Y|G=G_S)$
}
\label{pge_2}    
\end{equation}

PGExplainer determines $G_S$ by maximizing the mutual information $MI$ between the predictions $Y$ and this underlying structure with the help of entropy term $H$ as in \ref{pge_2}. The goal of the PGExplainer approach is to specifically explain the graph topologies found in GNNs and identify $G_S$ such that conditional entropy is minimized. However, because there are so many candidate values for $G_S$, direct optimization is infeasible. Therefore, assuming $G_S$ follows a Gilbert random graph distribution, selections of edges from the original input graph $G$ are conditionally independent of each other, a relaxation approach is used. With this relaxation, PGExplainer can recast the aim as an expectation, making optimization more manageable.

\begin{table*}[ht!]
\caption{Comparative Performance Evaluation of Edge Embedding Baselines}
\renewcommand{\arraystretch}{1.2}
\label{tab:cont4}
\centering
\resizebox{.8\textwidth}{!}{
\begin{tabular}{lccccccccc}
\hline
 & \multicolumn{3}{c}{NF-UNSW-NB15-v2} &  & \multicolumn{3}{c}{NF-CSE-CIC-IDS2018-v2} &  & Average across datasets \\ \hline
\textbf{}  & \textbf{F1-Macro} & \textbf{Acc} & \textbf{DR} & \textbf{} & \textbf{F1-Macro} & \textbf{Acc} & \textbf{DR} & \textbf{} & \textbf{F1-Macro} \\ \hline
Anomal-E 

& \textbf{92.35\%} & \textbf{98.66\%} & \textbf{98.77\%} &  
& 94.38\% & 97.80\% & 82.67\% & 
& \textbf{93.36\%} \\ \hline

DGI & 48.99\% & 96.02\%  & 0.00\% &  
& 46.82\% & 88.03\%  & 0.00\% &  
& 47.90\% \\ \hline

GraphSAGE & 54.77\% & 88.60\% & 26.63\% &  
& \textbf{94.61\%} & \textbf{97.90\%} & \textbf{82.60\%} &  
& 74.69\% \\ \hline
\end{tabular}
}
\end{table*}

\subsection{X-CBA: Explainability Aided CatBoosted Anomal-E}

The flow of the \textit{Explainability Aided CatBoosted Anomal-E} framework is presented in Fig. \ref{fig_proposed}. In the proposed framework, computer intrusion detection tabular data is transformed into attributed multigraph data after the required preprocessing steps are completed. Here, routers represent nodes and data flows represent edges. The self-learning DGI model is tuned with an E-GraphSAGE encoder and model training is performed on multigraph intrusion detection data. Gradient descent optimization, powered by the Adam Optimizer and the Binary Cross Entropy (BCE) Loss in \ref{eq_dgi}, is used to iteratively optimize the $D$, $R$, and $E$ of DGI. The encoder of the DGI uses a mean aggregation function on a 1-layer E-GraphSAGE model. E-GraphSAGE uses a hidden layer size of 256 units and ReLU is used as the activation function. As for the generation of the global graph summary, we averaged edge embeddings and passed them through a sigmoid function. BCE is used as a loss function and gradient descent is used for backpropagation with the Adam optimizer using a learning rate of 0.001. With the trained encoder, edge embeddings are obtained. Intrusion detection is performed with the CatBoost Classifier using the edge embeddings that outperform the meta-features of the data. The prediction results are explained with the help of PGExplainer, which is an XAI method designed specifically for graph data. In this way, the edges (data flows) that contribute the most to the prediction and are critical in the network topology are identified. The implementation details and the code repository of the proposed IDS are available here\footnote{https://github.com/kiymetkaya/xai-catboosted-anomale}.

\section{Performance Evaluation} \label{sec:eval}
We first evaluate the proposed X-CBA approach for intrusion detection performance comparing with the baselines in Section \ref{sec:eval-ids}. Then we also provide the explainability performance analysis with the state-of-the-art method in Section \ref{sec:eval-explainability}. 
\subsection{X-CBA Intrusion Detection Evaluation} \label{sec:eval-ids}
For the intrusion detection experiments, we have chosen baselines from two categories: (i) Step 5 in Modeling and (ii) Step 6 in Post-Modeling as shown in Fig. \ref{fig_proposed}. Here, Anomal-E, DGI, and GraphSAGE models are considered as edge embedding baselines for the evaluation of the X-CBA model from the Step 5 category. For the second category from Step 6, the classifier models to be used as baselines are as follows: 

\par \textbf{Principal Component Analysis (PCA) } algorithm adapted for intrusion detection, using a correlation matrix from ``benign" samples to identify attacks based on deviation from the benign correlation.

\textbf{Isolation Forest (IF)} utilizes tree structures to isolate attacks: samples closer to the root of the tree are identified as attacks, while the deeper ones in the tree are considered ``benign", creating an ensemble of trees for efficient and effective intrusion detection.

\textbf{Clustering-Based Local Outlier Factor (CBLOF)} treats intrusion detection as a clustering-based problem, assigning outlier factors based on cluster size and distance between a sample and its nearest cluster to determine if it is an attack.

\textbf{Histogram-Based Outlier Score (HBOS)} Employing a histogram-based approach, constructs univariate histograms for each feature and calculates bin densities, using these density values to determine the HBOS score and classify samples as attacks or not.

\textbf{AutoEncoder (AE)} is an unsupervised deep learning model, consisting of four \textit{Linear} layers, the first two used for encoding and the last two for decoding, which respectively transform \{number of features, 16, 8, 16, number of features\}. After each \textit{Linear} layer, ReLU is used as an activation function.

\textbf{Random Forest Classifier (RFC)} is a bootstrap ensemble model. It creates several decision trees on data samples and then selects the best solution using voting. RFC is chosen rather than a single decision tree because it reduces over-fitting and has less variance.

\textbf{XGBoost Classifier (XGBC)} stands for Extreme Gradient Boosting. XGBoost is an implementation of gradient boosting. The key idea behind the XGBC is the improvement of speed and performance using the reasons behind the good performance such as regularization and handling sparse data.

\textbf{LightGBM Classifier} is a histogram-based gradient boosting algorithm that reduces the computational cost by converting continuous variables into discrete ones. Since the training time of the decision trees is directly proportional to the number of computations and hence the number of splits, LightGBM provides a shorter model training time and efficient resource utilization.

The results presented in Table \ref{tab:cont4} are a summary result table for the prediction results of the Anomal-E, DGI, and GraphSAGE with the best corresponding baseline classifier model evaluated with the list above. `Macro Average F1-score' which is the average F1 score of all classes, `Accuracy' and `Detection Rate' (DR, also known as Recall) that measures the percentage of actual attack observations that the model correctly classified are chosen as performance evaluation metrics. As it can be seen from the results in Table \ref{tab:cont4}, the superiority of the Anomal-E over the state-of-the-art methods DGI and GraphSAGE has been proven with the experiments \cite{caville2022anomal} on NF-UNSW-NB15-v2 and NF-CSE-CIC-IDS2018-v2 datasets. 

After the preliminary experiments, in the second step, the proposed X-CBA approach is compared with Anomal-E \cite{caville2022anomal} since Anomal-E is the edge embedding approach among state-of-the-art as shown in Table \ref{tab:cont4}. For further detailed analysis, X-CBA and Anomal-E are evaluated with various classifier baselines located in Step 6 in the Post-Modeling. According to the results in Table \ref{tab:res}, where the prediction results are presented in ascending order according to the F1-Macro, the proposed X-CBA produced more accurate results comparing to Anomal-E in both unsupervised (with AutoEncoder) and supervised (with CatBoost) approaches among baseline models. In other words, Anomal-E utilizes edge embeddings and predicts attacks using IF, HBOS, PCA, and CBLOF methods together. The reason for this is that any one of these methods alone does not produce the best results in terms of all performance evaluation metrics. 

\begin{table}[ht]
\caption{Network Intrusion Detection Prediction Results}
\renewcommand{\arraystretch}{1.2}
\label{tab:res}
\centering
\resizebox{.4\textwidth}{!}{
\begin{tabular}{lllll}
\hline
 &  & \multicolumn{3}{c}{NF-CSE-CIC-IDS2018-v2} \\ \hline
\textbf{} &  & \multicolumn{1}{c}{\textbf{F1-Macro}} & \multicolumn{1}{c}{\textbf{Acc}} & \multicolumn{1}{c}{\textbf{DR}} \\ \hline
Anomal-E - IF &   & 81.11\% & 89.79\% & 91.84\% \\ \hline
Anomal-E - HBOS &   & 91.89\% & 96.86\% & 77.79\% \\ \hline
Anomal-E - PCA &  & 92.57\% & 97.11\%  & 79.16\% \\ \hline
Anomal-E - CBLOF &  & 94.38\%  & 97.80\% & 82.67\% \\ \hline \hline
X-CBA - RFC &  & 96.53\%  & 98.61\% &88.50\%  \\ \hline
X-CBA - AutoEncoder &  & 97.76\% & 99.13\% &94.92\%  \\ \hline
X-CBA - XGBC &  & 98.45\% & 99.36\% &94.85\%  \\ \hline
X-CBA - LightGBM &  & 98.56\% & 99.40\% &95.19\%  \\ \hline
\cellcolor[HTML]{C3E2C3}\textbf{X-CBA - CatBoost} & \cellcolor[HTML]{C3E2C3}& \cellcolor[HTML]{C3E2C3}\textbf{98.73\%}  & \cellcolor[HTML]{C3E2C3}\textbf{99.47\%} & \cellcolor[HTML]{C3E2C3}\textbf{95.74\%}  \\ \hline
\end{tabular}
}
\end{table}

On the other hand, when we utilize only CatBoost in the prediction phase with the X-CBA approach, we obtain the most accurate predictions with a single boosting model as seen in Table \ref{tab:res} green-backgrounded row. Table \ref{tab:res} presents the results we obtained on the NF-CSE-CIC-IDS2018-v2 dataset with ETC, RFC, AE, XGBC, LightGBM, and CatBoost ensemble models against the prediction algorithms in Anomal-E. NF-CSE-CIC-IDS2018-v2 dataset contains 18,893,708 network flows: 16,635,567 (88.05\%) benign samples and 2,258,141 (11.95\%) attack samples. For a fair comparison, we follow the same preprocessing steps, training procedures, and train-test split (70\% training, 30\% testing) as in Anomal-E \cite{caville2022anomal}. We then present our results using the same test set for consistency. Moreover, the best CatBoost model was determined with Scikit-learn GridSearch 5-fold-CV, just like the other ensemble models. Accordingly, the best hyperparameters for CatBoost are: ``min. samples split": 4, ``min. samples leaf": 4, ``max. depth": 8. 

\begin{figure*}[ht!]
    \centering
    \includegraphics[width=.8\linewidth]{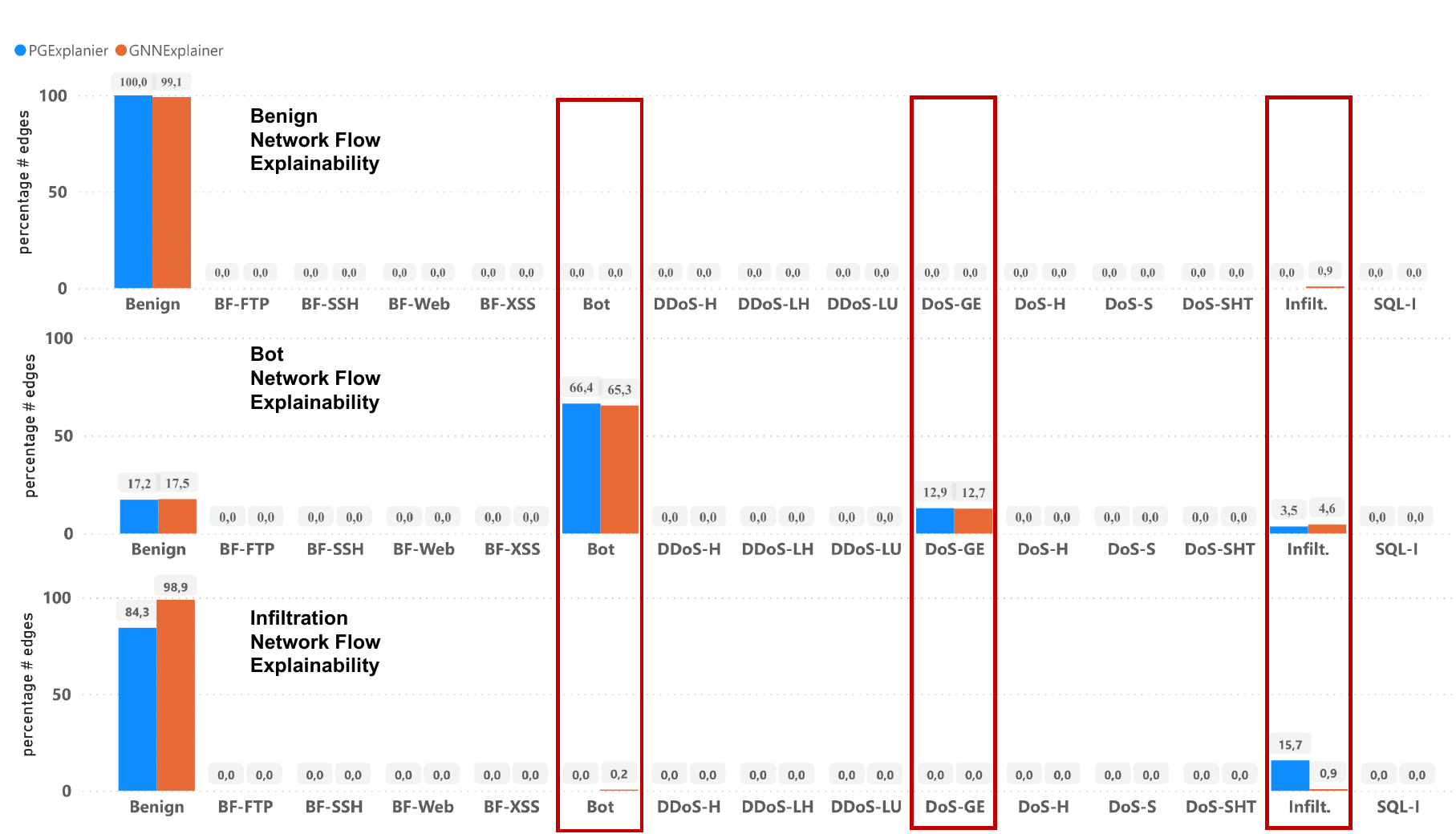}
    \caption{Distribution of edges according to attack type within the edge explanation maximizing mutual information for `Benign', `Bot' and `Infiltration'}
    \label{fig:xai}
\end{figure*}

\subsection{X-CBA Explainability Evaluation} \label{sec:eval-explainability}

The explainability performance of X-CBA approach implemented with PGExplainer \cite{luo2020parameterized} is evaluated through state-of-the-art XAI approaches. We analyze XAI methods designed specifically for GNNs that have the ability to explain important edges. Table \ref{tab_xai_comp} provides a summary of this analysis on XAI methods. In Table \ref{tab_xai_comp}, ``GNN Design" indicates whether the XAI method has a specific design for GNN models; ``Black Box" indicates whether the prediction model is treated as a black-box when being explained, and ``Target": represents the target component (N: nodes, NF: node features, E: edges.) whose explanation is presented. As can be seen in Table \ref{tab_xai_comp}, the explainability of GNNs' edges with the current state-of-the-art is measured by GNNExplainer and PGExplainer approaches. Other GNN explainability approaches are either focuses on nodes or node features, which is not suitable for \textit{flow-based network} intrusion detection explainability. 

\begin{table}[ht]
\caption{XAI Methods Comparison due to ``GNN Design", ``Black Box" and ``Target". }
\renewcommand{\arraystretch}{1.2}
\label{tab_xai_comp}
\centering
\resizebox{.45\textwidth}{!}{
\begin{tabular}{lccc}
\hline
Method & \multicolumn{1}{l}{GNN Design} & \multicolumn{1}{l}{Black Box} & \multicolumn{1}{l}{Target} \\ \hline
\textbf{CAM, Grad-CAM \cite{pope2019explainability}} & - & \textbf{-} & \textbf{N} \\ \hline
\textbf{LRP \cite{baldassarre2019explainability}} & \textbf{-} & \textbf{-} & \textbf{N} \\ \hline
\textbf{SHAP \cite{lundberg2017unified}} & \textbf{-} & \textbf{+} & \textbf{NF} \\ \hline
\textbf{GraphLIME\cite{huang2022graphlime}} & \textbf{+} & \textbf{+} & \textbf{NF} \\ \hline
\textbf{PGM-Explainer \cite{vu2020pgm}} & \textbf{+} & \textbf{+} & \textbf{N} \\ \hline
\textbf{ZORRO \cite{funke2022z}} & \textbf{+} & \textbf{+} & \textbf{N} \\ \hline
\textbf{\cellcolor[HTML]{F3F2C5}GNNExplainer \cite{ying2019gnnexplainer}} & \cellcolor[HTML]{F3F2C5}\textbf{+} &\cellcolor[HTML]{F3F2C5}\textbf{+} & \cellcolor[HTML]{F3F2C5}\textbf{E} \\ \hline
\color{blue} \textbf{X-CBA - PGExplainer \cite{luo2020parameterized}} & \textbf{\textcolor{blue}{+}} & \textbf{\textcolor{blue}{-}} & \textbf {\textcolor{blue}{ E}} \color{black} \\ \hline
\end{tabular}
}
\end{table}

Moreover, we evaluated the performance of network-flow importance (a.k.a edge importance) with two metrics \cite{yuan2022explainability} as used in edge explainability performance evaluation: $\text{Sparsity}$ metric in \ref{eq_sparsity} and $\text{Fidelity}+{}^{F}$ metric in \ref{eq_fidel}. The Sparsity whose formula is given in \ref{eq_sparsity} measures the proportion of important components (N, NF, E) identified by the XAI method. Here, ${\left| {m_i} \right|}$ shows the number of important edges determined by the XAI method, ${\left| {M_i} \right|}$ shows the number of all edges in the graph and $K$ is the number of graphs. On the other hand, $\text{Fidelity}+{}^{F}$ whose formula is given in \ref{eq_fidel} studies the change of prediction score where $F$ shows performance evaluation function (F1-Macro, Accuracy, and DR) \cite{yuan2022explainability}. ${G_i}^{1-m_{i}}$ represents the new graph obtained by keeping features of ${G_i}$ based on the complementary mask (${1-m_{i}}$)  and ${y_i}$ is the original prediction of the GNN model. 

\begin{figure}
    \centering
    \includegraphics[width=.82\linewidth]{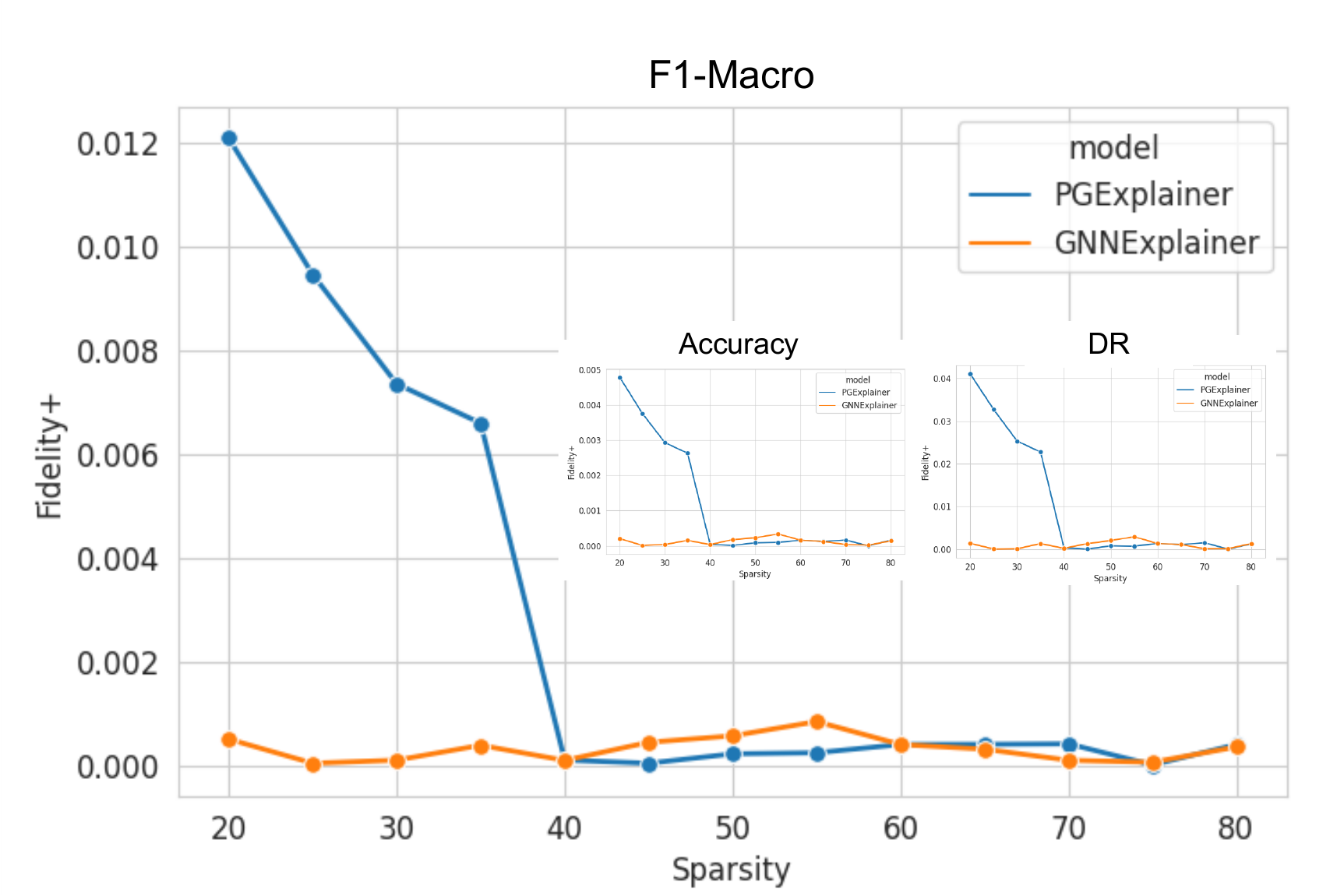}
    \caption{The $\text{Fidelity}+{}^{F}$ comparisons over F1-Macro, Accuracy and DR under different Sparsity levels}
    \label{fig_fidelityplus}
\end{figure}

\begin{equation}
\label{eq_sparsity}
\begin{aligned}
\text{Sparsity} = \frac{1}{K}\sum_{i=1}^{K}\left(1-\frac{\left| {m_i} \right|}{\left| {M_i} \right|}\right)
\end{aligned}
\end{equation}

\begin{equation}
\label{eq_fidel}
\text {Fidelity}+{}^F=\frac{1}{K} \sum_1^K\left(F\left(G_i\right)_{y_i}-F\left(G_i^{1-m_i}\right)_{y_i}\right)
\end{equation}

Fig. \ref{fig_fidelityplus} presents the \textit{global} graph's edge explanation results for PGExplainer and GNNExplainer with $\text{Fidelity}+{}^{F}$ and $\text{Sparsity}$. $\text{Fidelity}+{}^{F}$ measures the F1-Macro, Accuracy, and DR drops when important edges are removed. Higher $\text{Fidelity}+{}^{F}$ scores indicate the identified edges are more important for the proposed X-CBA. According to the results in Fig. \ref{fig_fidelityplus}, as it is expected, the PGExplainer outperforms the GNNExplainer significantly and consistently. This is because PGExplainer's non-black box explanation ability reveals the explanation for flow-based intrusion packets better.

Moreover, we investigate the most influential subgraphs that affect the prediction of an edge of a given attack type to \textit{locally} observe the differences between PGExplainer and GNNExplainer. NF-CSE-CIC-IDS2018-v2 includes data for benign and fifteen different attack types. For each attack type, we make use of the edges found to maximize mutual information by each XAI model. In Fig. \ref{fig:xai}, we illustrate the distribution of edge types in the subgraphs identified for three selected classes: Benign, Bot, and Infiltration. These classes were chosen from a total of 16 (15 attack classes and one benign class) to provide a diverse range of examples for analysis: Benign represents a non-attack scenario, Infiltration serves as a sample attack case, and Bot exemplifies a sophisticated attack. The prevalence of benign edges within subgraphs of various attacks is unsurprising, given the inherent dataset imbalance. The subgraph for `Benign' in Fig. \ref{fig:xai} is therefore expected to contain only benign edges, which is provided only by the PGExplainer, and failed in GNNExplainer. 

In contrast, Infiltration and Bot attacks are marked by a predominance of similar connection types in their immediate network environments. These attacks are known for their ability to spread across the network, often utilizing similar connections. In Fig. \ref{fig:xai}, illustrated by red rectangular boxes, PGExplainer distinguishes itself in identifying key features of both Infiltration and Bot attacks. However, its performance advantage is somewhat less pronounced for Bot attacks. This is because Bot attacks in our dataset are consistently linked to a specific node (IPV4 ADDR: 18.219.211.138), a pattern easily detected by both XAI methods. Consequently, the performance of the two methods is similar for Bot attacks.

PGExplainer's true strength is demonstrated in its ability to identify Infiltration flows. These flows are often challenging to detect as they are not always directly connected to the Infiltration attack. On the other hand, GNNExplainer significantly underperforms in this area, frequently misidentifying Infiltration attack instances as Bot attacks and missing most of the real Infiltration network flows. This discrepancy can be attributed to PGExplainer's operation in inductive settings and its capability to collectively explain multiple instances, enabling it to more effectively uncover local network flow relationships.

\section{Conclusion and Future Works} \label{sec:conc}
In this study, we propose a novel IDS methodology, called X-CBA, that synergizes the strengths of Graph Neural Networks (GNNs) and Explainable AI (XAI). X-CBA not only outperforms in detecting a wide array of cyber threats through its use of network flows and graph edge embeddings but also marks a significant leap in the accuracy and reliability of threat detection, as evidenced by its remarkable 99.47\% accuracy, 98.73\% F1 rate, and 95.74\% recall. Most importantly, X-CBA addresses the critical issue of transparency in ML/DL-based security solutions. We evaluated the baseline XAI methods to show the strong explainability of our proposed framework in terms of its ability to find important edges. By integrating PGExplainer, it provides both local and global explanations of its decision-making process and gives much more accurate results in terms of sparsity and fidelity metrics compared to baselines, enhancing trust and accountability in its operations.

\section*{Acknowledgements}
This research is supported by the Scientific and Technological Research Council of Turkey (TUBITAK) 1515 Frontier R\&D Laboratories Support Program for BTS Advanced AI Hub: BTS Autonomous Networks and Data Innovation Lab. project number 5239903, TUBITAK 1501 project number 3220892, and the ITU Scientific Research Projects Fund under grant number MÇAP-2022-43823.

\bibliographystyle{IEEEtran}
\bibliography{bibliography/bibliography}
\end{document}